\documentclass{PoS}

\usepackage{amsmath}

\newcommand{\Ref}[1]{(\ref{#1})}
\newcommand{\MC}[1]{\ensuremath{\mathcal{#1}}}

\newcommand{\Kahler}{K\"ahler}

\newcommand{\RA}{\ensuremath{\rightarrow}}

\newcommand{\C}{\ensuremath{\mathbb{C}}}
\newcommand{\ot}{\ensuremath{\otimes}}
\newcommand{\compos}{\ensuremath{\circ}}
\newcommand{\del}{\ensuremath{\partial}}
\newcommand{\vp}{\ensuremath{\varphi}}
\newcommand{\stat}[1]{%
	{\ensuremath{|#1|}}}
\newcommand{\act}{\ensuremath{\operatorname{\rhd}}}

\newcommand{\Qt}{\ensuremath{\tilde{Q}}}
\newcommand{\phit}{\ensuremath{\tilde{\phi}}}
\newcommand{\psit}{\ensuremath{\tilde{\psi}}}
\newcommand{\sigmat}{\ensuremath{\tilde{\sigma}}}
\newcommand{\muhat}{\ensuremath{\hat{\mu}}}
\newcommand{\statfctr}{\ensuremath{(-1)^{\mathcal{F}}}}

\newcommand{\product}{\ensuremath{m}}
\newcommand{\fproduct}{\ensuremath{\cdot}}
\newcommand{\opproduct}{\ensuremath{\cdot}}

\newcommand{\coproduct}{\ensuremath{\Delta}}
\newcommand{\counit}{\ensuremath{\epsilon}}
\newcommand{\antipode}{\ensuremath{\operatorname{S}}}
\newcommand{\transpose}{\ensuremath{\operatorname{\tau}}}
\newcommand{\id}{\ensuremath{\operatorname{id}}}
\newcommand{\vid}{%
 {\settowidth\unitlength{\ensuremath{\boldsymbol{1}}}%
  \begin{picture}(1,1)
  \put(0.385,0){\makebox[0pt]{\ensuremath{\mathrm{1}}}}
  \put(0.615,0){\makebox[0pt]{\ensuremath{\mathrm{l}}}}
  \end{picture}}}
\newcommand{\fid}{\ensuremath{\mathbf{1}}}
\newcommand{\braid}{\ensuremath{\Psi}}

\title{Lattice supersymmetry with Hopf algebra for the link approach}

\ShortTitle{Lattice supersymmetry with Hopf algebra for the link approach}

\author{A.~D'Adda,\\
        Dipartimento di Fisica Teorica, Universita' di Torino
								and INFN Sezione di Torino\\
								I-10125 Torino, Italy\\
        E-mail: \email{dadda@to.infn.it}}
\author{N.~Kawamoto and \speaker{J.~Saito}\\
        Department of Physics, Hokkaido University\\
								Sapporo, 060-0810 Japan\\
        E-mail:
        \email{kawamoto@particle.sci.hokudai.ac.jp},\\
        E-mail:
								\email{saito@particle.sci.hokudai.ac.jp}}

\abstract{%
A formalism of lattice supersymmetry
based on a lattice-deformed superalgebra which
was originally introduced in the link approach
formulation is presented.
We propose that the superalgebra can in fact be
identified as a Hopf algebra, showing
all the Hopf algebra axioms and consistencies
are satisfied with explicit formulae.
In particular, the ``deformed'' Leibniz rules
proposed in the original link approach are now built in
the coproduct structure of the Hopf algebra.
Fields in this scheme, as representations of the Hopf
algebra, are found to obey a kind of mildly 
deformed statistics, which is interpreted
as a braiding strucutre. We can then construct,
at least perturbatively,
the corresponding lattice field theory, which
has the Hopf algebraic symmetry with the deformed
statistics, as an example
of braided quantumm field theory formulated
by Oeckl.}

\FullConference{The XXVII International Symposium on Lattice Field Theory\\
                July 26-31, 2009\\
                Peking University, Beijing, China}

\begin{document}

\section{Introduction}

Regularization of supersymmetric field theory has increased
its importance both for recent theoretical and phenomenological
developments, including, for instance,
gauge/gravity duals such as AdS/CFT and supersymmetry breaking.
One may expect that
lattice formulation, among others, would provide a promising
regularization scheme with the applicability to strong-coupling,
thus constructive and nonperturbative, analysis in the first
principle calculation.
It is, however, far from straightforward to incorporate
supersymmetry on the lattice due to the discrete nature of
spacetime;
superalgebra, which prescribes supersymmetry,
contains momentum operator, and momentum operator should be
the generator of infinitesimal spacetime translation,
which is broken on the lattice.
For this difficulty, various approaches have been
developed so far. (For a review
see \cite{review} and references therein.)
In this article, we present a possible formulation~\cite{DKS} of lattice
supersymmetry with a "deformed" notion of superalgebra
in the framework of the link approach~\cite{DKKN, ADFKS}.
This deformation can be naturally interpreted as a generalization
of Lie algebra to Hopf algebra. What we need to formulate is then
a field theory with this Hopf algebraic supersymmetry. We show
that such a formulation would be given by applying a general
formalism called braided quantum field theory (BQFT)~\cite{Oeckl}.
For this
purpose, we introduce a simply generalized statistics of fields
which is compatible with the structure of our Hopf algebra.
Supersymmetry on the lattice can now be recognized as various
sets of Ward--Takahashi identities derived by this BQFT formalism
\cite{Sasai-Sasakura}.

We will illustrate these aspects in the following,
mainly concentrating on two-dimensional non-gauge examples.

\section{Superalgebra in the link approach}

In the link approach~\cite{DKKN},
superalgebra on a two-dimensional lattice
was introduced in the form
\begin{equation}
	\{Q, Q_\mu\} = i\del_{+\mu},\qquad
	\{\Qt, Q_\mu\} = -i\epsilon_{\mu\nu}\del_{-\nu},
\label{t-algebra}
\end{equation}
with the other commutators just vanishing.
Notice that supercharges $Q_A=Q,\ Q_\mu$
and $\Qt$
are expressed in
the Dirac--\Kahler\ twisted basis, which essentially corresponds
to $\MC{N}=(2,2)$ supercharges in two dimensions in the normal
basis.%
\footnote{With a similar argument, we need to take
the Dirac--\Kahler\ twisted $\MC{N}=4$ supersymmetry in
four dimensions.}
We can see that fermions in the link approach
should be geometrically distributed on the lattice
just like the Dirac--\Kahler\
or staggered fermions, where the d.o.f.\ of possible
doublers on the lattice is essentially used as that of
extended supersymmetry through the Dirac--\Kahler\ twisting. This is why
the twisted basis was chosen in the superalgebra above.
Another point is that the algebra \Ref{t-algebra}
contains the forward and backward finite difference
operators $\del_{\pm\mu}$ which simply replace the momentum
operator in the continuum. These difference operators
don't obey the Leibniz rule, but obey the modified Leibniz rule
\begin{equation}
	\del_{\pm\mu}(\vp_1\fproduct\vp_2)(x)
	=
	\del_{\pm\mu}\vp_1(x)\vp_2(x)
	+
	\vp_1(x\pm a\muhat)\del_{\pm\mu}\vp_2(x),
\label{mod-Leib-P}
\end{equation}
where $\muhat$ denotes the unit vector along the $\mu$ direction.
One might expect that, other less simple operators,
instead of the simple forward/backward difference operators,
could obey the usual Leibniz rule even on the lattice.
This is not possible, however, due to the no-go theorem
proving non-existence of such a local operator \cite{Kato-Sakamoto-So},
which implies the modified Leibniz rule is unavoidable on the lattice
when one concentrate on local field theories.
In other words, it implies that the supercharges
can't obey the normal Leibniz rule either to make the algebra
\Ref{t-algebra} hold.

In the link approach,
the following modified Leibniz rule for $Q_A$ was assumed
\begin{equation}
	Q_A(\vp_1\cdot\vp_2)(x)
	=
	Q_A\vp_1(x)\vp_2(x)
	+(-1)^\stat{\vp_1}\vp_1(x+a_A)Q_A\vp_2(x),
\label{mod-Leib-Q}
\end{equation}
where $\stat{\vp}$ is 0 (or 1) when
$\vp$ is bosonic (or fermionic, respectively).
This already shows that the algebra \Ref{t-algebra}
doesn't form a Lie superalgebra in the usual sense,
and ``deforms'' the notion of usual superalgebra.
Then a natural question is whether we could treat
the algebra \Ref{t-algebra}
with the modified Leibniz rules \Ref{mod-Leib-P} and \Ref{mod-Leib-Q}
in a mathematically rigorous manner.
We will see shortly that the answer is affirmative;
the algebra in the link approach can be identified as a Hopf algebra,
which assures mathematical consistency especially
for the modified Leibniz rule \Ref{mod-Leib-Q}.
Another question would immediately
follows: even if the algebra itself makes sense, it is still
unclear whether it corresponds to a symmetry of a local quantum field
theory as usual Lie algebra does. For this, too, we will propose
an affirmative answer.
First,
we can manage to take care of the locality with a mildly generalized
statistics of fields. This statistics is in fact expressed
mathematically as (trivial) braiding.
Since quantum field theory for fields with such a braiding structure
is known to be formulated generally as BQFT~\cite{Oeckl},
we could apply it to our case.
This approach now allows us to associate our Hopf algebraic symmetry
with various sets of Ward--Takahashi identities~\cite{Sasai-Sasakura},
showing clear relations of the Hopf algebra to a symmetry
of a quantum field theory.

\section{Superalgebra on the lattice as a Hopf algebra}

Here we are going to show how the superalgebra which was originally
introduced with an extra shift structure in the link approach~\cite{DKKN}
can as a whole be rigorously identified as a Hopf algebra \cite{DKS}.
Before going into the detail,
let us briefly summarize the Hopf algebra axioms.
For a full mathematical treatment on Hopf algebra, see, for example,
\cite{Majid}.

Hopf algebra $H$ is an object which satisfies the following
four axioms.
\begin{enumerate}

\item $H$ is an algebra, namely a vector space which has
an associative product (multiplication)
$\opproduct: H\ot H\RA H$, where the associativity reads
$\opproduct\compos(\opproduct\ot\id)=\opproduct\compos(\id\ot\opproduct)$,
and unit element 
$\vid$.\footnote{Here $\id$ is the identity map and $\compos$
denotes composition of maps.}

\item $H$ is a coalgebra, namely a vector space which has
a coassociative coproduct (comultiplication)
$\coproduct: H\RA H\ot H$, where the coassociativity reads
\begin{equation}
	(\coproduct\ot\id)\compos\coproduct=(\id\ot\coproduct)\compos\coproduct,
\label{coassociativity}
\end{equation}
and counit $\counit: H\RA\C$ which satisfies the condition
\begin{equation}
	(\counit\ot\id)\compos\coproduct=(\id\ot\counit)\compos\coproduct=\id.
\label{counit-condition}
\end{equation}

\item The coproduct and counit are both algebra maps, namely,
\begin{equation}
	\begin{cases}
	\coproduct\compos\opproduct = (\opproduct\ot\opproduct)\compos\coproduct,
	\\
	\coproduct(\vid) = \vid\ot\vid,
	\end{cases}
	\qquad\text{and}\qquad
	\begin{cases}
	\counit\compos\opproduct = \counit\ot\counit,\\
	\counit(\vid)=1.
	\end{cases}
\label{algebra-maps}
\end{equation}

\item $H$ has an antipode $\antipode: H\RA H$, which satisfies
the defining condition
\begin{equation}
	\opproduct\compos(\antipode\ot\id)\compos\coproduct
	=\opproduct\compos(\id\ot\antipode)\compos\coproduct
	=\vid\counit.
\label{antipode-condition}
\end{equation}

\end{enumerate}

It is easy to see that
the superalgebra~\Ref{t-algebra} in the link approach
forms an algebra. Product of two
generators, say, $Q_A$ and $Q_B$, is defined with a
successive applications of $Q_B$ and $Q_A$ as in
\(
	(Q_A\opproduct Q_B)\act\vp
	:= (Q_A\act)\compos(Q_B\act)\vp,
\)%
\footnote{The ``action'' of a generator $a$ on a field $\vp$
is denoted as $a\act\vp$.}
whereas the unit operator is trivially defined as
$\vid\act\vp=\vp$.
These structures together with the ``equivalence'' relations,
(anti-)commutation relations \Ref{t-algebra}, form
a universal enveloping algebra of a sort.
To be specific, let us list explicit field representations
for this algebra, taking the example of
$\MC{N}=(2,2)$ Wess--Zumino model in two dimensions.
The field contents are
scalar bosons $\phi,\ \sigma$, fermions
$\psi,\ \psi_\mu,\ \psit$ and auxiliary fields
$\phit,\ \sigmat$,
for which the supertransformations are as follows:
\begin{equation}
	\begin{aligned}
	Q\phi &= 0, &\qquad Q_\mu\phi &= \psi_\mu, &\qquad \Qt\phi &= 0, \\
	Q\psi_\nu &= i\del_{+\nu}\phi, &\qquad
	Q_\mu\psi_\nu &= -\epsilon_{\mu\nu}\phit, &\qquad
	\Qt\psi_\nu &= -i\epsilon_{\nu\mu}\del_{-\mu}\phi, \\
	Q\phit &= -i\epsilon_{\mu\nu}\del_{+\mu}\psi_\nu, &\qquad
	Q_\mu\phit &= 0, &\qquad
	\Qt\phit &= i\del_{-\mu}\psi_\mu, \\
	Q\sigma &= -\psi, &\qquad Q_\mu\sigma &= 0, &\qquad
	\Qt\sigma &= -\psit, \\
	Q\psi &= 0, &\qquad Q_\mu\psi &= -i\del_{+\mu}\sigma, &\qquad
	\Qt\psi &= -\sigmat, \\
	Q\psit &= \sigmat, &\qquad
	Q_\mu\psit &= i\epsilon_{\mu\nu}\del_{-\nu}\sigma, &\qquad
	\Qt\psit &= 0, \\
	Q\sigmat &= 0, &\qquad
	Q_\mu\sigmat &=
	i\epsilon_{\mu\nu}\del_{-\nu}\psi +i\del_{+\mu}\psit,
	&\qquad
	\Qt\sigmat &= 0.
	\end{aligned}
\label{supertransf}
\end{equation}

What is more important is the coproduct structure. It
amounts to specifying the action of an operator, say $Q_A$,
on a product of fields
$\vp_1\fproduct\vp_2 =: \product(\vp_1\ot\vp_2)$ as in
\begin{equation}
	Q_A\act(\vp_1\fproduct\vp_2)
	= \product\Bigl(
	\coproduct(Q_A)\act(\vp_1\ot\vp_2)
	\Bigr).
\label{act-on-prod}
\end{equation}
Thus determining the coproduct structure
is nothing but to specifying the Leibniz rule.
For instance,
the modified Leibniz rule \Ref{mod-Leib-Q} is essentially equivalent
to the coproduct formula
\begin{equation}
	\coproduct(Q_A)
	= Q_A\ot\vid + \statfctr\opproduct T_{a_A}\ot Q_A,
\label{coproduct-Q}
\end{equation}
where $\statfctr$ just gives factor $+1$ (or $-1$)
when applied on a bosonic (or fermionic, resp.)\ field:
$\statfctr\act\vp = \pm\vp$,
and $T_{a_A}$ is the shift operator:
$\bigl(T_{a_A}\act\vp\bigr)(x) := \vp(x+a_A)$.
Note in passing that these operators would satisfy
the trivial (anti-)commutation relations
\begin{equation}
		[Q_A, T_b] = [P_\mu, T_b] = [T_b, T_c] = 0,\qquad
	\{Q_A,\statfctr\}
	= [P_A,\statfctr]
	= [T_b,\statfctr]
	= 0.
\end{equation}
We can determine the coproduct for
$\del_{\pm\mu},\ T_{b}$ and $\statfctr$ by the identifications
similar to \Ref{act-on-prod}, which result in the following
formulae:
\begin{equation}
		\coproduct(\del_{\pm\mu})
		= \del_{\pm\mu}\ot\vid + T_{\pm a\muhat}\ot\del_{\pm\mu},\qquad
		\coproduct(T_b) = T_b\ot T_b,\qquad
		\coproduct(\statfctr) = \statfctr\ot\statfctr.
\label{coproduct-others}
\end{equation}
We can confirm ourselves, by straightforward calculations,
that these prescriptions indeed obey the coassociativity
condition~\Ref{coassociativity}.
Notice that the coassociativity condition
assures the uniqueness of the action of an operator
on a product of three or more fields.
For instance, since by the associativity
\(
\vp_1\fproduct\vp_2\fproduct\vp_3
=
\bigl(\vp_1\fproduct\vp_2\bigr)\fproduct\vp_3
=
\vp_1\fproduct\bigl(\vp_2\fproduct\vp_3\bigr),
\)
we need
\(
Q_A\act\bigl(\vp_1\fproduct\vp_2\fproduct\vp_3\bigr)
=
Q_A\act\Bigl(\bigl(\vp_1\fproduct\vp_2\bigr)\fproduct\vp_3\Bigr)
=
Q_A\act\Bigl(\vp_1\fproduct\bigl(\vp_2\fproduct\vp_3\bigr)\Bigr).
\)
This is equivalent to
\(
(\coproduct\ot\id)\compos\coproduct(Q_A)
=
(\id\ot\coproduct)\compos\coproduct(Q_A),
\)
which is the coassociativity condition for the operator $Q_A$.
Similar arguments of course hold for other operators.

The coproduct structure thus determines how operators act on
products of fields.
Note, however, that any field $\vp$ can be considered as
a product
\(
	\vp = \fid\fproduct\vp = \vp\fproduct\fid
\)
with the constant field $\fid$. Accordingly,
when the operator, say, $Q_A$
acts on $\vp$, it must satisfy the consistency condition
\(
		Q_A\act\vp
		= Q_A\act\bigl(\fid\fproduct\vp)
		= Q_A\act\bigl(\vp\fproduct\fid).
\)
In order to state this more generally, let us define
the counit map by
\begin{equation}
	Q_A\act\fid
	\equiv
	\counit(Q_A)\fid,
\end{equation}
so that the counit gives the trivial representation.
The consistency above is now written as
\(
		\id
		= (\counit\ot\id)\compos\coproduct = (\id\ot\counit)\compos\coproduct,
\)
which is what we listed before \Ref{counit-condition}.
The explicit formulae \Ref{coproduct-Q} and \Ref{coproduct-others}
now allow us to specify the counit of operators which satisfies
this condition as follows:
\begin{equation}
			\counit(Q_A) = 0,\quad
			\counit(P_\mu) = 0,\quad
			\counit(T_b) = 1,\quad
			\counit\bigl(\statfctr\bigr) = 1. 
\label{counit}
\end{equation}

Coproduct and counit for a product of operators
can be calculated through the algebra-map conditions
\Ref{algebra-maps}. We emphasize that this property
is important since it also assures
the explicit formulae
\Ref{coproduct-Q}, \Ref{coproduct-others} and
\Ref{counit} are indeed compatible with
the algebraic relations \Ref{t-algebra}.

We introduce one more object, the antipode.
It essentially gives the ``inverse'' of operators,
uniquely determined by the relation \Ref{antipode-condition}.
From the explicit formulae \Ref{coproduct-Q},
\Ref{coproduct-others} and \Ref{counit},
we find the following formulae
\begin{equation}
		\antipode(Q_A) 
		=
		-\statfctr\opproduct Q_A,
		\qquad
		\antipode(P_\mu) 
		= -P_\mu,\qquad
		\antipode(T_b) = T_b^{-1},\qquad
		\antipode\bigl(\statfctr\bigr) = \statfctr. 
\label{antipode}
\end{equation}
We can show by the condition \Ref{antipode-condition}
that the antipode is anti-algebraic, namely,
\(
	\antipode\compos\opproduct
	=
	\opproduct\compos\transpose\compos
	(\antipode\ot\antipode)
\)
and
\(
	\antipode(\vid) = \vid,
\)
where $\transpose$ is the transposition
$\transpose(a\ot b) := b\ot a$.
This is again consistent with the relation \Ref{t-algebra},
as seen with the explicit formulae
\Ref{coproduct-Q}, \Ref{coproduct-others} and \Ref{counit}.
We can also derive that the antipode should satisfy
anti-coalgebraic nature as in
\(
	(\antipode\ot\antipode)\compos\coproduct
	=
	\transpose\compos\coproduct\compos\antipode
\)
and
\(
	\counit\compos\antipode
	=
	\counit,
\)
which are also found to be compatible to the explicit formulae.

\section{Statistics on the lattice as a braiding}

Our next task is to consider field-product representations of the Hopf
algebraic supersymmetry.
We first emphasize here that a Hopf algebra in general has
a noncommutative representation. In the current application,
a noncommutative representation would naturally
lead to a noncommutative field theory, which would then be
nonlocal. In fact, we could avoid this noncommutativity
or nonlocality, systematically taking product representations which are
almost commutative, or, in other words, commutative
up to a mildly generalized statistics.%
\footnote{Our Hopf algebra has a simple structure,
(quasi)triangularity, which allows such an almost
commutative representation.}

We illustrate more concretely
how this is possible with the previous example of
$\MC{N}=(2,2)$ Wess--Zumino model in two dimensions.
For scalars $\phi,\ \sigma$, supertransformations
with respect to $Q_\mu$ are given as
$Q_\mu\phi=\psi_\mu,\ Q_\mu\sigma=0$ (see \Ref{supertransf}).
Let us assume that the scalars be commutative:
$\phi(x)\fproduct\sigma(x) = \sigma(x)\fproduct\phi(x)$.
The point is, once we set this assumption, we could deduce
from the supertransformations \Ref{supertransf} the statistics
for the other fields in a manner totally consistent with
the Hopf algebra structures. To see this, we calculate
$Q_\mu\bigl(\phi(x)\fproduct\sigma(x)\bigr)
=Q_\mu\bigl(\sigma(x)\fproduct\phi(x)\bigr)$,
so that, with the use of the coproduct formula \Ref{coproduct-Q},
we have
\(
	\psi_\mu(x)\fproduct\sigma(x)
	=
	\sigma(x+a_\mu)\fproduct\psi_\mu(x).
\)
This shows that the fermion $\psi_\mu$ is commutative with
the boson $\sigma$ up to the shift of argument. Similar
calculations lead to that $\psi_\mu$ is
(anti-)commutative with any other fields up to the same
amount of shift of argument. We can in fact generalize this
statement as follows
\begin{equation}
	\braid\Bigl(
	\vp_{A_0\cdots A_p}(x)\ot\vp'_{B_0\cdots B_q}(y)
	\Bigr)
	=
	(-1)^{pq}
	\vp'_{B_0\cdots B_q}\left(y+\sum_{i=1}^p a_{A_i}\right)
	\ot \vp_{A_0\cdots A_p}\left(x-\sum_{i=1}^q a_{B_i}\right),
\label{braid}
\end{equation}
where
$\braid$ represents the exchange of the order of fields
in a tensor product, called (trivial) braid, and
\(
\vp_{A_0\cdots A_p}:= Q_{A_p}\cdots Q_{A_1}\vp_{A_0},
\)
with $\vp_{A_0}$ denoting scalars $\phi$ or $\sigma$.

%

With the mildly generalized statistics \Ref{braid},
the ordering ambiguity claimed in \cite{Dutch} no longer appears.%
\footnote{Another difficulty raised there
in the case of gauge theory needs further investigations.}
Notice also that we might understand this statistics property
in terms of a grading structure for each field
and symmetry operator which
is determined corresponding to the indices $A_i$ and
$B_i$ in the formula \Ref{braid}.
Then in particular the difference operators $\del_{\pm\mu}$ as well
must have the grading structure, which is difficult to express
explicitly. We will come to this point in the conclusion.

\section{Supersymmetric lattice field theory as BQFT}

Quantum field theory for fields with generalized statistics,
or braiding, can be generally formulated as braided quantum
field theory (BQFT) at least perturbatively~\cite{Oeckl}.
We can thus apply the formalism into our approach
to construct a perturbative lattice field theory.
Here we just sketch the outline of the formulation.
The theory is quantized through path integral formalism
\begin{equation}
	Z = \int e^{-S},\qquad
	\langle\vp_1\cdots\vp_n\rangle
	=\frac{1}{Z}\int\vp_1\cdots\vp_n e^{-S},\qquad
		\int\frac{\delta}{\delta\vp(x)} = 0,
\end{equation}
for a classical action $S$. The last equation
formally defines the path integral, for which
the functional derivative is assumed to obey
the deformed Leibniz rule
\begin{equation}
		\frac{\delta}{\delta\vp(x)} (\vp_1\fproduct\vp_2)
		= \frac{\delta}{\delta\vp(x)}\vp_1\fproduct\vp_2
		+ (-1)^{\stat{\vp}\stat{\vp_1}}T_\vp^{-1}\vp_1\fproduct
				\frac{\delta}{\delta\vp(x)}\vp_2.
\end{equation}
The formal expression is enough to derive perturbative
Wick's theorem with appropriate statistics, which allows
to compute arbitrary correlation functions in terms of propagators
determined by the specific form of the classical action.
Now the classical Hopf algebraic supersymmetry is expressed
by $Q_A\act S = 0$. At the quantum level, this leads
to various sets of Ward--Takahashi identities of the form 
\cite{Sasai-Sasakura}
\begin{equation}
		Q_A\act\langle\vp_1\cdots\vp_n\rangle
		=
		\counit(Q_A)\langle\vp_1\cdots\vp_n\rangle
		= 0.
\end{equation}

\section{Conclusion}

In this article, we presented a formulation of lattice
supersymmetry	with the machinery of Hopf algebra and
BQFT, based on the previously proposed formulation,
the link approach.
We showed explicitly that superalgebra on a lattice can be
identified as a Hopf algebra, where the modified or
deformed Leibniz rules invented in the original link
approach are now incorporated as the coproduct structure
in the Hopf algebra. Fields, as representations of
the Hopf algebraic symmetry, would in general be noncommutative,
thus the corresponding field theory would be nonlocal.
This noncommutativity, however, could be reduced to
the commutativity up to a lattice-deformed statistics
in a manner consistent with the Hopf algebraic superalgebra.
The difficulty claimed as ordering ambiguity
against the original link approach
is now solved thanks to this deformed statistics.
We then applied the formalism of BQFT to construct
a quantum field theory for such a generalized statistics,
which allows us to derive Ward--Takahashi identities
associated with the Hopf algebraic supersymmetry
at least perturbatively.

In this formulation, fields and symmetry generators
could be interpreted to have grading structures corresponding to
the deformed statistics of fields.
In particular, ``momentum'' operator,
the difference operators on the lattice, should have nontrivial grading.
In order to compute arbitrary correlation functions,
especially when including loop corrections,
we need an explicit representation of the graded
difference operators. Such a representation might be
unnecessary for the computations of physical observables.
Another issue is that this construction at the moment is limited
only on a formal and perturbative level, and it is not
yet clear whether it can lead to nonperturbative
formulation as a lattice field theory.
Gauge theory extension is missing as well.
These issues are for the future works.



\end{document}